\RequirePackage[mathlines]{lineno} 
\documentclass[amsmath,amssymb,twocolumn,prd,floatfix,showpacs,nofootinbib]{revtex4-1}
\usepackage{overpic,graphicx}
\usepackage{dcolumn}
\usepackage{bm}
\usepackage{rotating}
\usepackage{subfigure}
\usepackage{color}
\usepackage{epstopdf}

\usepackage{ulem}
\usepackage{enumerate}
\begin{document}

\title{\bf \boldmath
Observation of the Semileptonic $D^+$ Decay into the $\bar K_1(1270)^0$ Axial-Vector Meson
}

\author{M.~Ablikim$^{1}$, M.~N.~Achasov$^{10,d}$, P.~Adlarson$^{59}$, S. ~Ahmed$^{15}$, M.~Albrecht$^{4}$, M.~Alekseev$^{58A,58C}$, A.~Amoroso$^{58A,58C}$, F.~F.~An$^{1}$, Q.~An$^{55,43}$, Y.~Bai$^{42}$
% ...
\\
\vspace{0.2cm}
(BESIII Collaboration)\\
\vspace{0.2cm} {\it
$^{1}$ Institute of High Energy Physics, Beijing 100049, People's Republic of China\\
% ...
$^{4}$ Bochum Ruhr-University, D-44780 Bochum, Germany\\
% ...
$^{10}$ G.I. Budker Institute of Nuclear Physics SB RAS (BINP), Novosibirsk 630090, Russia\\
% ...
$^{15}$ Helmholtz Institute Mainz, Johann-Joachim-Becher-Weg 45, D-55099 Mainz, Germany\\
% ...
$^{42}$ Southeast University, Nanjing 211100, People's Republic of China\\
$^{43}$ State Key Laboratory of Particle Detection and Electronics, Beijing 100049, Hefei 230026, People's Republic of China\\
% ...
$^{55}$ University of Science and Technology of China, Hefei 230026, People's Republic of China\\
% ...
$^{58}$ (A)University of Turin, I-10125, Turin, Italy; (B)University of Eastern Piedmont, I-15121, Alessandria, Italy; (C)INFN, I-10125, Turin, Italy\\
$^{59}$ Uppsala University, Box 516, SE-75120 Uppsala, Sweden\\
% ...
$^{d}$ Also at the Novosibirsk State University, Novosibirsk, 630090, Russia\\
% ...
}
}

\begin{abstract}
By analyzing a 2.93~$\rm fb^{-1}$ data sample of $e^+e^-$ collisions,
recorded at a center-of-mass energy of 3.773 $\rm \,GeV$ with the BESIII detector operated at the BEPCII collider,
we report the first observation of the semileptonic $D^+$ transition into the axial-vector meson $D^{+} \rightarrow {\bar{K}}_{1}(1270)^{0}e^{+}\nu_{e}$
with a statistical significance greater than $10\sigma$.
Its decay branching fraction is determined to be
${\mathcal B}[D^+\to \bar K_1(1270)^0 e^+\nu_e]=(2.30\pm0.26^{+0.18}_{-0.21}
\pm 0.25)\times10^{-3}$, where the first and second uncertainties are
statistical and systematic, respectively, and the third originates from the input branching
fraction of ${\bar{K}}_{1}(1270)^{0} \rightarrow K^{-} \pi^{+} \pi^{0}$.
\end{abstract}


\maketitle

\renewcommand{\thefootnote}{\fnsymbol{footnote}}
\footnotetext{Corresponding author: liuke@ihep.ac.cn}

Studies of semileptonic (SL) $D$ transitions,
mediated via $c\to s(d)\ell^+\nu_\ell$ at the quark level,
are important for the understanding of nonperturbative strong-interaction dynamics in weak decays~\cite{isgw,isgw2}.
Those transitions into S-wave states have been extensively studied in theory and experiment.
However, there is still no experimental confirmation of the predicted transitions into P-wave states.

In the quark model, the physical mass eigenstates of the strange axial-vector mesons, $K_1(1270)$ and $K_1(1400)$, are mixtures of the
$^1\rm P_1$ and $^3\rm P_1$ states with a mixing angle $\theta_{K_1}$.
These mesons have been thoroughly studied via $\tau$, $B$, $D$, $\psi(3686)$ and $J/\psi$ decays, as well as via
$Kp$ scattering~\cite{4,5,6,7,8,14,jpsi,chic0,1,2}.
Nevertheless, the value of $\theta_{K_1}$ is still very controversial in various phenomenological
analyses~\cite{18,19,20,21,22,24,26,27}.
Studies of the SL $D$ transitions into $\bar K_1(1270)$
provide important insight into the mixing angle $\theta_{K_1}$.
The improved knowledge of $\theta_{K_1}$ is essential for theoretical calculations describing the decays of $\tau$~\cite{18}, $B$~\cite{20,B-K1g},
and $D$~\cite{D-AP,D-AP-1} particles into strange axial-vector
mesons, and for investigations in the field of hadron spectroscopy~\cite{Burakovsky}.

Earlier quantitative predictions for the branching fractions (BFs) of $D^{0(+)}\to \bar K_{1}(1270) e^+ \nu_e$ were derived from
the Isgur-Scora-Grinstein-Wise (ISGW) quark model~\cite{isgw} and its update, ISGW2~\cite{isgw2}.
ISGW2 implies that the BFs of $D^{0(+)}\to \bar K_{1}(1270) e^+ \nu_e$ are about 0.1\,(0.3)\%. However, the model ignores the
mixing between $^1\rm P_1$ and $^3\rm P_1$ states.
Recently, the rates of these decays were calculated with three-point QCD sum rules (3PSR)~\cite{Khosravi},
covariant light-front quark model (CLFQM)~\cite{theory}, and light-cone QCD sum rules (LCSR)~\cite{formfactor}.
In general, the predicted BFs range from $10^{-3}$ to $10^{-2}$~\cite{Khosravi,theory,formfactor}, and are
sensitive to $\theta_{K_1}$ and its sign.
Measurements of $D^{0(+)}\to \bar K_1(1270)e^+\nu_e$ will be critical to
distinguish between theoretical calculations, to explore the nature of strange axial-vector mesons,
and to understand the weak-decay mechanisms of $D$ mesons.

Currently, there is very little experimental information available about semileptonic $D$ decays
into axial-vector mesons, with the only result being the reported
evidence for the process $D^0 \to K_1(1270)^- e^+\nu_e$ from the CLEO Collaboration~\cite{cleores}.
This Letter presents the first observation of $D^+\to \bar K_1(1270)^0e^+\nu_e$~\cite{charge}
by using an $e^+e^-$ data sample corresponding to an integrated luminosity of 2.93~fb$^{-1}$~\cite{lum} recorded at a
center-of-mass energy of $\sqrt s=3.773$ $\rm \,GeV$ with the BESIII detector~\cite{Ablikim:2009aa}.

Details about the design and performance of the BESIII detector are given in Ref.~\cite{Ablikim:2009aa}.
Simulated samples produced with the {\sc geant4}-based~\cite{geant4} Monte Carlo (MC) package, which
includes the geometric description of the BESIII detector and the
detector response, are used to determine the detection efficiency
and to estimate the backgrounds. The simulation includes the beam-energy spread and initial-state radiation (ISR) in the $e^+e^-$
annihilations modeled with the generator {\sc kkmc}~\cite{ref:kkmc}.
The inclusive MC samples consist of the production of the $D\bar{D}$
pairs, the non-$D\bar{D}$ decays of the $\psi(3770)$, the ISR
production of the $J/\psi$ and $\psi(3686)$ states, and the
continuum processes incorporated in {\sc kkmc}~\cite{ref:kkmc}.
The known decay modes are modeled with {\sc
evtgen}~\cite{ref:evtgen} using BFs taken from the
Particle Data Group~\cite{pdg2018}, and the remaining unknown decays
from the charmonium states with {\sc
lundcharm}~\cite{ref:lundcharm}. The final-state radiation (FSR)
from charged final-state particles are incorporated with the {\sc
photos} package~\cite{photos}.
The $D^+\to \bar K_1(1270)^0e^+\nu_e$ decay is simulated with
the ISGW2 model~\cite{MPM}, the $\bar K_1(1270)^0$ is set to
decay into all possible processes containing the $K^-\pi^+\pi^0$ combination.
The resonance shape of $\bar K_1(1270)^0$ is parameterized by a relativistic Breit-Wigner function, and the mass and width of $\bar K_1(1270)^0$ are fixed at the world-average values 1272$\pm$7 MeV and 90$\pm$20 MeV, respectively~\cite{pdg2018}.

The measurement employs the $e^+e^-\to \psi(3770)\to D^+D^-$ decay chain.
The $D^-$ mesons are reconstructed by their hadronic decays to $K^{+}\pi^{-}\pi^{-}$,
$K^0_{S}\pi^{-}$, $K^{+}\pi^{-}\pi^{-}\pi^{0}$, $K^0_{S}\pi^{-}\pi^{0}$, $K^0_{S}\pi^{+}\pi^{-}\pi^{-}$,
and $K^{+}K^{-}\pi^{-}$. These inclusively selected events are referred to as single-tag (ST) $D^-$ mesons.
In the presence of the ST $D^-$ mesons, candidate $D^+\to \bar K_{1}(1270)^0e^+ \nu_{e}$ decays
are selected to form double-tag (DT) events.
The BF of $D^+\to \bar K_{1}(1270)^0 e^+ \nu_{e}$ is given by

\begin{equation}
\label{eq:bf}
{\mathcal B}_{\rm SL}=N_{\mathrm{DT}}/(N_{\mathrm{ST}}^{\rm tot}\cdot \varepsilon_{\rm SL}),
\end{equation}
where $N_{\rm ST}^{\rm tot}$ and $N_{\rm DT}$ are the ST and DT yields in the data sample,
$\varepsilon_{\rm SL}=\Sigma_i [(\varepsilon^i_{\rm DT}\cdot N^i_{\rm ST})/(\varepsilon^i_{\rm ST}\cdot N^{\rm tot}_{\rm ST})]$ is the efficiency of detecting the SL decay in the presence of the ST $D^-$ meson.
Here $i$ denotes the tag mode, and $\varepsilon_{\rm ST}$ and
$\varepsilon_{\rm DT}$ are the ST and DT efficiencies of selecting the ST and DT candidates, respectively.

We use the same selection criteria as discussed in Refs.~\cite{epjc76_369,cpc40_113001,prl121_171803}.
All charged tracks are required to be within a polar-angle ($\theta$) range of $|\rm{cos\theta}|<0.93$.
All of them, except for those from $K^0_{S}$ decays, must originate from an interaction region defined by
$V_{xy}<$ 1~cm and $|V_{z}|<$ 10~cm. Here, $V_{xy}$ and $|V_{z}|$ denote the distances of closest approach
of the reconstructed track to the interaction point (IP) in the $xy$ plane and the $z$ direction (along the beam), respectively.

Particle identification (PID) of charged kaons and pions is performed using
the specific ionization energy loss ($dE/dx$) measured by the main drift chamber (MDC) and the time-of-flight.
Positron PID also uses the measured information from the electromagnetic calorimeter (EMC). The combined confidence levels
under the positron, pion, and kaon hypotheses ($CL_e$, $CL_{\pi}$ and $CL_{K}$, respectively) are calculated.
Kaon (pion) candidates are required to satisfy $CL_{K}>CL_{\pi}$ ($CL_{\pi}>CL_{K}$). Positron
candidates are required to satisfy $CL_e>0.001$ and $CL_e/(CL_e+CL_\pi+CL_K)>0.8$. To reduce the background
from hadrons and muons, the positron candidate is further required to have a deposited energy in the EMC greater than 0.8 times its momentum in the MDC.

$K^0_{S}$ candidates are reconstructed from two oppositely charged tracks satisfying $|V_{z}|<$ 20~cm.
The two charged tracks are assigned
as $\pi^+\pi^-$ without imposing further PID criteria. They are constrained to
originate from a common vertex and are required to have an invariant mass
within $|M_{\pi^{+}\pi^{-}} - M_{K_{S}^{0}}|<$ 12~MeV$/c^{2}$, where
$M_{K_{S}^{0}}$ is the $K^0_{S}$ nominal mass~\cite{pdg2018}. The
decay length of the $K^0_S$ candidate is required to be greater than
twice the vertex resolution away from the IP.

Photon candidates are selected using the information from the EMC.
It is required that the shower time is within 700~ns of the event start time,
the shower energy be greater than 25 (50)~MeV
if the crystal with the maximum deposited energy in that cluster
is in the barrel (end-cap) region~\cite{Ablikim:2009aa},
and the opening angle between the candidate shower and
any charged tracks is greater than $10^{\circ}$.
Neutral $\pi^0$ candidates are selected from the photon pairs with the invariant mass within $(0.115, 0.150)$ GeV$/c^{2}$.
The momentum resolution of the accepted photon pair is improved by a kinematic fit, which
constrains the $\gamma\gamma$ invariant mass to the $\pi^{0}$ nominal mass~\cite{pdg2018}.

The ST $D^{-}$ mesons are distinguished from the combinatorial backgrounds by two variables:
the energy difference $\Delta E=E_{D^-} - E_{\rm beam}$ and
the beam-energy constrained mass $M_{\rm BC} = \sqrt{E^{2}_{\rm beam}-|\vec{p}_{D^-}|^{2}}$,
where $E_{\rm beam}$ is the beam energy,
and $\vec{p}_{D^-}$ and $E_{D^-}$ are the measured momentum and energy of
the ST candidate in the $e^+e^-$ center-of-mass frame, respectively.
For each tag mode,
only the one with the minimum $|\Delta E|$ is kept.
The combinatorial backgrounds in the $M_{\rm BC}$ distributions are suppressed by requiring
$\Delta E$ within $(-55, +40)$~MeV for the tag modes involving a $\pi^0$,
and $(-25, +25)$~MeV for the other tag modes.

\begin{figure}[htp]
  \centering
  \includegraphics[width=1.0\linewidth]{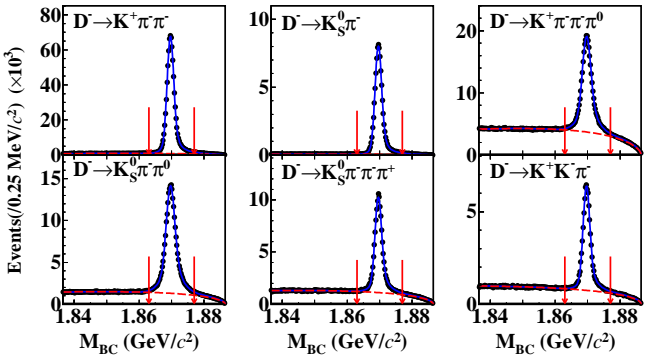}
  \caption{
The $M_{\rm BC}$ distributions of the ST candidates in the data sample (dots with error bars).
Blue solid curves are the fit results and
red dashed curves represent the background contributions of the fit.
The pair of red arrows in each subfigure indicate the $M_{\rm BC}$ window.
  }\label{fig:datafit_Massbc}
\end{figure}

Figure~\ref{fig:datafit_Massbc} shows the $M_{\rm BC}$ distributions of the accepted ST candidates in the data sample
for various tag modes.
The ST yield for each tag mode is obtained by performing a maximum-likelihood fit to the corresponding $M_{\rm BC}$ distribution.
In the fits,
the $D^{-}$ signal is modeled by an MC-simulated $M_{\rm BC}$ shape
convolved with a double-Gaussian function
and the combinatorial-background shape is described by an
ARGUS function \cite{ARGUS}.
The candidates in the $M_{\rm BC}$ signal region, $(1.863,1.877)$~GeV/$c^2$,
are kept for further analysis.
The total ST yield is $N^{\rm tot}_{\rm ST}= 1522474 \pm 2215$, where the uncertainty is statistical.

In the analysis of the particles recoiling against the ST $D^-$ mesons,
candidate events for the $D^+\to \bar K_{1}(1270)^0 e^+ \nu_{e}$ channel are selected from the remaining tracks that
have not been used for the ST reconstruction. The $\bar K_{1}(1270)^0$ meson is reconstructed using its
dominant decay $\bar K_{1}(1270)^0\to K^- \pi^+\pi^0$. It is required that there are only three good charged tracks available for this selection.
One of the tracks with charge opposite to that of the $D^-$ tag is identified as the positron.
The other two oppositely charged tracks are identified as a kaon and a pion, according to their PID information.
Moreover, the kaon candidate must have charge opposite to that of the positron.
Other selection criteria, which have been optimized by analyzing the inclusive MC samples, are  as follows.
To effectively veto the backgrounds associated with wrongly paired photons,
the $\pi^0$ candidates must have a momentum greater than 0.15\,GeV/$c$ and a decay angle
$|\cos\theta_{\rm decay,\pi^0}|=|E_{\gamma_1}- E_{\gamma_2}|/|\vec{p}_{\pi^0}|$ less than 0.8.
Here, $E_{\gamma_1}$ and $E_{\gamma_2}$ are the energies of $\gamma_1$ and $\gamma_2$, and $\vec{p}_{\pi^0}$
is the momentum of the $\pi^0$ candidate.
To suppress the potential backgrounds from the hadronic decays $D^+\to K^-\pi^+\pi^+\pi^0$,
the invariant mass of the $K^-\pi^+\pi^0 e^+$ combination, $M_{K^-\pi^+\pi^0 e^+}$, is required to
be smaller than 1.78~GeV/$c^2$.

Information concerning the undetectable neutrino is inferred by the kinematic quantity
$U_{\mathrm{miss}}\equiv E_{\mathrm{miss}}-|\vec{p}_{\mathrm{miss}}|$,
where $E_{\mathrm{miss}}$ and $\vec{p}_{\mathrm{miss}}$ are the missing energy and momentum
of the SL candidate, respectively,
calculated by $E_{\mathrm{miss}}\equiv E_{\mathrm{beam}}-\Sigma_j E_j$
and $\vec{p}_{\mathrm{miss}}\equiv\vec{p}_{D^{+}}-\Sigma_j \vec{p}_j$
in the $e^+e^-$ center-of-mass frame. The index $j$ sums over the $K^-$, $\pi^+$, $\pi^0$ and $e^+$ of the signal candidate,
and $E_j$ and $\vec{p}_j$ are the energy and momentum of the $j$th particle, respectively.
To improve the $U_{\mathrm{miss}}$ resolution, the
$D^+$ energy is constrained to the beam energy and $\vec{p}_{D^{+}}
\equiv -\hat{p}_{D^-}\sqrt{E_{\mathrm{beam}}^{2}-m_{D^{+}}^{2}}$, where
$\hat{p}_{D^-}$ is the unit vector in the momentum direction of
the ST $D^{-}$, and $m_{D^{+}}$ is the $D^+$ nominal
mass~\cite{pdg2018}.
To partially recover the effects of FSR and bremsstrahlung (FSR recovery), the four-momenta of photon(s) within
$5^\circ$ of the initial positron direction are added to the positron four-momentum measured by the MDC.

Events that originate from the process $D^+\to \bar K^{*}(892)^0[\to K^-\pi^+]e^+\nu_{e}$,
in which a fake $\pi^0$ is wrongly associated to the signal decay,
form a peaking background around $+0.02$ GeV in the $U_{\rm miss}$ distribution
and around 1.15~GeV/$c^2$ in the $M_{K^-\pi^+\pi^0}$ distribution.
To suppress these backgrounds, we define an alternative kinematic quantity
$U_{\mathrm{miss}}^\prime \equiv E_{\mathrm{miss}}^\prime - |\vec{p}_{\mathrm{miss}}^\prime|$,
where $E_{\mathrm{miss}}^\prime \equiv  E_{\mathrm{beam}} - \Sigma_j E_j$ and
$\vec{p}_{\mathrm{miss}}^\prime \equiv  \vec{p}_{D^+} - \Sigma_j \vec{p}_j$,
and $j$ only sums over the $K^-$, $\pi^+$ and $e^+$ candidates of the signal candidate.
Since these backgrounds form an obvious peak around zero in the $U_{\mathrm{miss}}^\prime$ distribution,
the $U_{\rm miss}^\prime$ values of the SL candidates are required to lie outside $(-0.09, 0.03)$\,GeV.

\begin{figure*}[htbp]\centering
  \includegraphics[width=1.0\linewidth]{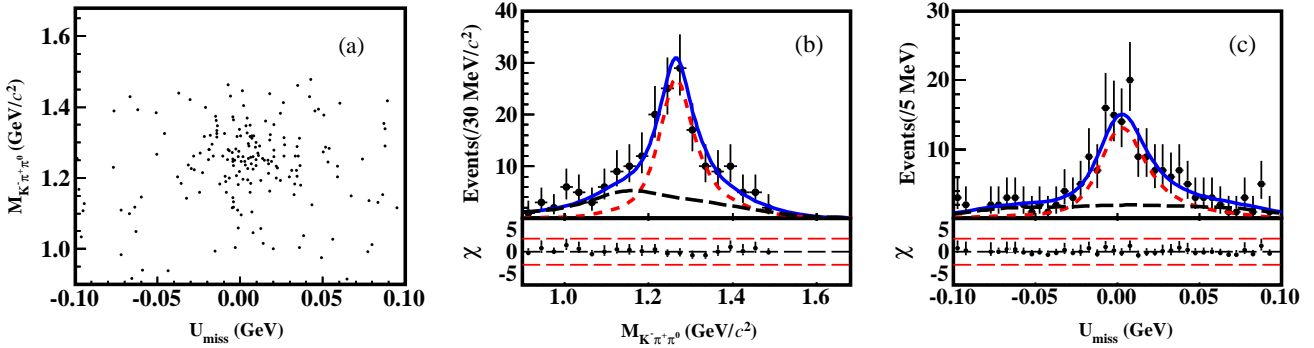}
  \vspace*{-0.5cm}
\caption{
(a) The $M_{K^-\pi^+\pi^0}$ vs. $U_{\rm miss}$ distribution of the SL candidate events and
(b,\,c) the projections to $M_{K^-\pi^+\pi^0}$ and $U_{\rm miss}$, respectively, with the residual $\chi$ distributions of the 2-D fit.
Dots with error bars are data.
Blue solid, red and black dashed curves are the fit result,
the fitted signal and the fitted background, respectively.
}
\label{fig:fit}
\end{figure*}

Figure~\ref{fig:fit}\,(a) shows the distribution of $M_{K^-\pi^+\pi^0}$ vs. $U_{\rm miss}$ of the accepted
$D^+\to K^-\pi^+\pi^0e^+\nu_e$ candidate events in the data sample after combining all tag modes.
A clear signal, which concentrates around 1.27~GeV/$c^2$ in the $M_{K^-\pi^+\pi^0}$ distribution and around zero in the
$U_{\rm miss}$ distribution, can be seen.
The DT yield is obtained from a two-dimensional (2-D) unbinned extended maximum-likelihood fit
of the data presented by the distribution in Fig.~\ref{fig:fit}(a).  In the fit,
the 2-D signal shape is described by the MC-simulated shape extracted from the signal MC events of
$D^+\to\bar K_{1}(1270)^0 e^{+}\nu_{e}$.
The 2-D background shape is modeled by the MC-simulated shape
obtained from the inclusive MC samples and the number of background events is a free parameter in the fit.
The smooth 2-D probability density functions of signal and background are modeled by the corresponding MC-simulated shape~\cite{roofit,class}.
The projections of the 2-D fit on the $M_{K^-\pi^+\pi^0}$ and $U_{\rm miss}$ distributions
are shown in Figs.~\ref{fig:fit}\,(b) and~\ref{fig:fit}\,(c).
In the fit, we ignore the contributions from non-resonant decays
$D^+\to K^-\pi^+\pi^0 e^{+}\nu_{e}$, $\bar K^{*}(892)^0 \pi^0e^{+}\nu_{e}$, $K^{*}(892)^-\pi^+e^{+}\nu_{e}$ and
$K^-\rho(770)^+e^{+}\nu_{e}$, as well as the possible interference between them due to
the low significance of these contributions with the limited size of the data set.
The two decays $D^+\to \bar K_1(1400)^0 e^{+}\nu_{e}$ and $D^+\to \bar K^{*}(1430)^0 e^{+}\nu_{e}$ are indistinguishable, and as no significant contribution is found from either source, these components are not included in the fit.
From the fit, we obtain the DT yield of
$N_{\rm DT}=119.7\pm13.3$, where the uncertainty is statistical.
The statistical significance of the signal is estimated to be greater than 10$\sigma$, by comparing the likelihoods with and
without the signal components included, and taking the change in the number of degrees of freedom into account.

For each tag mode, the DT efficiency is estimated with the corresponding signal MC events.
The average signal efficiency is determined to be $\varepsilon_{\rm SL}=0.0742\pm0.0007$.
Compared to $\epsilon_{\rm SL}$, the signal efficiencies for individual tag modes vary within $\pm10\%$.
The reliability of the MC simulation is tested by examining typical distributions of the SL candidate events.
The data distributions of momenta and $\cos\theta$ of $K^-$, $\pi^+$, $\pi^0$ and $e^+$ are consistent with those of MC simulations.

By inserting $N_{\rm DT}$, $\varepsilon_{\rm SL}$, and $N_{\rm ST}^{\rm tot}$ into Eq.~(\ref{eq:bf}), we determine
the product of ${\mathcal B}_{\rm SL}$ and the BF of $\bar K_1(1270)^0\to
K^-\pi^+\pi^0$ (${\mathcal B}_{\rm sub}$) to be
$${\mathcal B}_{\rm SL}\cdot{\mathcal B}_{\rm sub}=(1.06\pm0.12^{+0.08}_{-0.10})\times10^{-3},$$
where the first and second uncertainties are statistical and systematic, respectively.

The systematic uncertainties in the BF measurement, which are assigned relative to the measured BF,
are discussed below.  The DT method ensures that most uncertainties arising from the ST selection cancel.
The uncertainty from the ST yield is assigned to be 0.5\%~\cite{epjc76_369,cpc40_113001,prl121_171803},
by examining the relative change in the yield between data and MC simulation after
varying the $M_{\rm BC}$ fit range, the signal shape, and the endpoint of the ARGUS function.

The uncertainties associated with the efficiencies of $e^+$ tracking (PID), $K^-$ tracking (PID), $\pi^+$ tracking (PID) and $\pi^0$ reconstruction
are investigated using data and MC samples of  $e^+e^-\to\gamma e^+ e^-$ events and DT $D\bar D$ hadronic events.
Small differences between the data and MC efficiencies are found, which are
$-(0.03\pm0.15)\%$, $+(0.94\pm0.27)\%$, $+(2.63\pm0.32)\%$, $-(0.14\pm0.18)\%$, $+(0.03\pm0.13)\%$, $-(0.08\pm0.18)\%$
for $e^+$ tracking, $e^+$ PID, $K^-$ tracking, $K^-$ PID, $\pi^+$ tracking and $\pi^+$ PID, respectively.
The MC efficiency is then corrected by these differences and used to determine the central value of the BF.
In the studies of $e^+$ tracking (PID) efficiencies,
the 2-D (momentum and $\cos\theta$) tracking efficiencies of data and MC simulation
of $e^+e^-\to\gamma e^+ e^-$ events are re-weighted to match those of $D^+ \to \bar K_1(1270)^0 e^+\nu_e$ decays.
After corrections, we assign the uncertainties associated with the $e^+$ tracking (PID), $K^-$ tracking (PID),
$\pi^+$ tracking (PID) and $\pi^0$ reconstruction
to be 1.0\%\,(1.0\%), 1.0\%\,(0.5\%), 0.5\%\,(0.5\%) and 2.0\%, respectively.

The uncertainty associated with  the $M_{K^-\pi^+\pi^0 e^+}$ requirement is estimated by varying
the requirement by $\pm0.05$ GeV/$c^2$, and
the largest change on the BF, 0.9\%, is taken as the systematic
uncertainty.
Similarly, the systematic uncertainty in the $U_{\rm miss}^{\prime}$ requirement is estimated
to be 1.7\% by varying the corresponding selection window by $\pm0.01$~GeV.
The uncertainty of the input BFs of $\bar K_1(1270)^0$ is estimated by changing the BF of each subdecay
by $\pm 1\sigma$.
The largest variation in the detection efficiency, 0.5\%, is assigned as the related systematic uncertainty.
The uncertainty of the 2-D fit is estimated
to be $^{+7.0\%}_{-8.2\%}$ by examining the BF changes with different fit ranges, signal shapes (dominated by varying the width of $\bar K_1(1270)^0$ by $\pm 1\sigma$) and background shapes. The uncertainty arising from background shapes is mainly due to unknown non-resonant decays, and is assigned as the change of the fitted DT yield when
they are fixed by referring to the well known non-resonant fraction in $D^+\to \bar K^*(892)^0e^+\nu_e$~\cite{non-resonance}.
The uncertainty arising from the limited size of the MC samples is 1.0\%.

The uncertainty due to FSR recovery is evaluated to be 1.3\%
which is the change of the BF when varying the FSR recovery angle to be $10^\circ$.
The total systematic uncertainty is estimated to be $^{+8.0\%}_{-9.0\%}$ by
adding all the individual contributions in quadrature.

When making use of  the world average of
${\mathcal B}_{\rm sub}=0.467\pm0.050$~\cite{pdg2018,K10},
we obtain 
$${\mathcal B}_{\rm SL}=(2.30\pm0.26^{+0.18}_{-0.21}
\pm 0.25)\times10^{-3},$$
where the third uncertainty, 10.7\%, is from the external uncertainty of the input BF ${\mathcal B}_{\rm sub}$.

To summarize, by analyzing an $e^+e^-$ collision data sample of 2.93~$\rm fb^{-1}$ taken at $\sqrt s=3.773$~GeV,
we report the observation of $D^+\to \bar K_1(1270)^0e^+\nu_e$ and determine its decay BF for the first time.
The measured BF is 1.4\% of the total semileptonic $D^+$ decay width, which lies between the ISGW prediction of 1\%
and the ISGW2 prediction of 2\%.
Our BF of $D^+\to \bar K_1(1270)^0 e^+\nu_e$ agrees with
the CLFQM and LCSR predictions when $\theta_{K_1}\approx 33^\circ$ or $57^\circ$~\cite{theory},
and clearly rules out the predictions when setting $\theta_{K_1}$ negative~\cite{formfactor}.
Making use of the measured value for the BF of $D^0\to K_1(1270)^-e^+\nu_e$~\cite{cleores} and the world-average lifetimes of the $D^0$ and $D^+$ mesons~\cite{pdg2018}, we determine the partial decay width ratio $\Gamma [D^+\to \bar K_1(1270)^0 e^+\nu_e]/\Gamma [D^0\to K_1(1270)^- e^+\nu_e]=1.2^{+0.7}_{-0.5}$,
which is consistent with unity as predicted by isospin conservation.
This demonstration  of the capability to observe $\bar K_1(1270)$ mesons in the very clean environment of SL $D^{0(+)}$ decays
opens up the opportunity to conduct further studies of the nature of these axial-vector mesons.
A near-future follow-up analysis of the dynamics of these SL decays with higher statistics will allow for
deeper explorations of the inner structure, production, mass and width of $\bar K_1(1270)$ and $\bar K_1(1400)$,
as well as providing access to hadronic-transition form factors.

The BESIII collaboration thanks the staff of BEPCII and the IHEP computing center for their strong support.
Authors thank helpful discussions from Xianwei Kang and Haiyang Cheng.
This work is supported in part by National Key Basic Research Program of China under Contract No. 2015CB856700; National Natural Science Foundation of China (NSFC) under Contract No. 11835012; National Natural Science Foundation of China (NSFC) under Contracts No. 11775230, No. 11625523, No. 11635010, No. 11735014; the Chinese Academy of Sciences (CAS) Large-Scale Scientific Facility Program; Joint Large-Scale Scientific Facility Funds of the NSFC and CAS under Contracts Nos. U1532257, U1532258, U1732263, U1832107, U1832207; CAS Key Research Program of Frontier Sciences under Contracts Nos. QYZDJ-SSW-SLH003, QYZDJ-SSW-SLH040; 100 Talents Program of CAS; INPAC and Shanghai Key Laboratory for Particle Physics and Cosmology; German Research Foundation DFG under Contract No. Collaborative Research Center CRC 1044, FOR 2359; Istituto Nazionale di Fisica Nucleare, Italy; Koninklijke Nederlandse Akademie van Wetenschappen (KNAW) under Contract No. 530-4CDP03; Ministry of Development of Turkey under Contract No. DPT2006K-120470; National Science and Technology fund; The Knut and Alice Wallenberg Foundation (Sweden) under Contract No. 2016.0157; The Royal Society, UK under Contract No. DH160214; The Swedish Research Council; U. S. Department of Energy under Contracts Nos. DE-FG02-05ER41374, DE-SC-0010118, DE-SC-0012069; University of Groningen (RuG) and the Helmholtzzentrum fuer Schwerionenforschung GmbH (GSI), Darmstadt.


\begin{thebibliography}{**}

\bibitem{isgw}
N. Isgur, D. Scora, B. Grinstein, and M. B. Wise, Phys. Rev. D {\bf 39}, 799 (1989).

\bibitem{isgw2}
D. Scora and N. Isgur, Phys. Rev. D {\bf 52}, 2783 (1995).

\bibitem{4} R. Barate {\it et al.} (ALEPH Collaboration), Eur. Phys. J. C {\bf 11}, 599 (1999).
\bibitem{5} G. Abbiendi {\it et al.} (OPAL Collaboration), Eur. Phys. J. C {\bf 13}, 197 (2000).
\bibitem{6} D. M. Asner {\it et al.} (CLEO Collaboration), Phys. Rev. D {\bf 62}, 072006 (2000).
\bibitem{7} K. Abe {\it et al.} (Belle Collaboration), Phys. Rev. Lett. {\bf 87}, 161601 (2001).
\bibitem{8} H. Yang {\it et al.} (Belle Collaboration), Phys. Rev. Lett. {\bf 94}, 111802 (2005).
\bibitem{14} J. M. Link {\it et al.} (FOCUS Collaboration), Phys. Lett. B {\bf 610}, 225 (2005).
\bibitem{jpsi} J. Z. Bai {\it et al.} (BES Collaboration),  Phys. Rev. Lett. {\bf 83}, 1918 (1999).
\bibitem{chic0} M.~Ablikim {\it et al.} (BES Collaboration), Phys. Rev. D {\bf 72}, 092002 (2005).
\bibitem{1} C. Daum {\it et al.} (ACCMOR Collaboration), Nucl. Phys. B {\bf 187}, 1 (1981).
\bibitem{2} D. Aston {\it et al.}, Nucl. Phys. B {\bf 292}, 693 (1987).
\bibitem{18} M. Suzuki, Phys. Rev. D {\bf 47}, 1252 (1993).
\bibitem{19} F. Divotgey, L. Olbrich and F. Giacosa, Eur. Phys. J. A {\bf 49}, 135 (2013).
\bibitem{20} H. Hatanaka and K. C. Yang, Phys. Rev. D {\bf 77}, 094023 (2008); Phys. Rev. D {\bf 78}, 059902(E) (2008).
\bibitem{21} H. Y. Cheng, Phys. Lett. B {\bf 707}, 116 (2012).
\bibitem{22} H. G. Blundell, S. Godfrey and B. Phelps, Phys. Rev. D {\bf 53}, 3712 (1996).
\bibitem{24} A. Tayduganov, E. Kou and A. Le Yaouanc, Phys. Rev. D {\bf 85}, 074011 (2012).
\bibitem{26} H. J. Lipkin, Phys. Lett. B {\bf 72}, 249 (1977).
\bibitem{27} L. Burakovsky and T. Goldman, Phys. Rev. D {\bf 56}, R1368 (1997).
\bibitem{B-K1g}
H. Y. Cheng and C. K. Chua, Phys. Rev. D {\bf 69}, 094007 (2004); Phys. Rev. D {\bf 81}, 059901(E) (2010).
\bibitem{D-AP}
H. Y. Cheng and C. W. Chiang, Phys. Rev. D {\bf 81}, 074031 (2010).
\bibitem{D-AP-1}
H. Y. Cheng, Phys. Rev. D {\bf 67}, 094007 (2003).
\bibitem{Burakovsky}
L. Burakovsky and T. Goldman, Phys. Rev. D {\bf 57}, 2879 (1998).
\bibitem{Khosravi}
R. Khosravi, K. Azizi, and N. Ghahramany, Phys. Rev. D {\bf 79}, 036004 (2009).
\bibitem{theory}
H. Y. Cheng and X. W. Kang, Eur. Phys. J. C {\bf 77}, 587 (2017);
Eur. Phys. J. C {\bf 77}, 863(E) (2017), and private communication.
\bibitem{formfactor}
S. Momeni and R. Khosravi, J. Phys. G {\bf 46}, 105006 (2019).
\bibitem{cleores}
M. Artuso {\it et al.} (CLEO Collaboration), Phys. Rev. Lett. {\bf 99}, 191801 (2007).
\bibitem{charge}
Throughout the Letter, charged conjugated modes are implied unless stated explicitly.
\bibitem{lum}
M.~Ablikim {\it et al.} (BESIII Collaboration),
Chin. Phys. C {\bf 37}, 123001 (2013);
Phys. Lett. B {\bf 753}, 629 (2016).
\bibitem{Ablikim:2009aa}
  M.~Ablikim {\it et al.} (BESIII Collaboration),
  Nucl.\ Instrum.\ Meth.\ A {\bf 614}, 345 (2010).
\bibitem{geant4}
  S.~Agostinelli {\it et al.} (GEANT4 Collaboration),
  Nucl.\ Instrum.\ Meth.\ A {\bf 506}, 250 (2003).
\bibitem{ref:kkmc}
  S.~Jadach, B.~F.~L.~Ward, and Z.~Was,
  Phys.\ Rev.\ D {\bf 63}, 113009 (2001);
  Comput.\ Phys.\ Commun.\  {\bf 130}, 260 (2000).
\bibitem{ref:evtgen}
  D.~J.~Lange,
  Nucl.\ Instrum.\ Meth.\ A {\bf 462}, 152 (2001);
  R.~G.~Ping,
  Chin. Phys. C {\bf 32}, 599 (2008).
\bibitem{pdg2018} M. Tanabashi {\it et al.} (Particle Data Group), Phys. Rev. D {\bf 98}, 030001 (2018) and 2019 update.


\bibitem{ref:lundcharm}
  J.~C.~Chen, G.~S.~Huang, X.~R.~Qi, D.~H.~Zhang, and Y.~S.~Zhu,
  Phys.\ Rev.\ D {\bf 62}, 034003 (2000);
  R.~L.~Yang, R.~G.~Ping, and H.~Chen,
  Chin.\ Phys.\ Lett.\  {\bf 31}, 061301 (2014).

\bibitem{photos}
  E.~Richter-Was,
  Phys.\ Lett.\ B {\bf 303}, 163 (1993).


\bibitem{MPM} D. Becirevic and A. B. Kaidalov, Phys. Lett. B {\bf 478}, 417 (2000).

\bibitem{epjc76_369}
M. Ablikim {\it et al}. (BESIII Collaboration), Eur. Phys. J. C {\bf 76}, 369 (2016).

\bibitem{cpc40_113001}
M. Ablikim {\it et al}. (BESIII Collaboration), Chin. Phys. C {\bf 40}, 113001 (2016).

\bibitem{prl121_171803}
M. Ablikim {\it et al}. (BESIII Collaboration), Phys. Rev. Lett. {\bf 121}, 171803 (2018).

\bibitem{ARGUS} H. Albrecht {\it et al.} (ARGUS Collaboration), Phys. Lett. B {\bf 241}, 278 (1990).
\bibitem{roofit} W. Verkerke and D. Kirkby, eConf No.C0303241 (2003)MOLT007[arXiv:physics/0306116].

\bibitem{class} https://root.cern.ch/doc/master/classRooNDKeysPdf.html.

\bibitem{non-resonance}
M. Ablikim {\it et al}. (BESIII Collaboration), Phys. Rev. D {\bf 94}, 032001 (2016).

\bibitem{K10}$\mathcal B_{K_1\to K^-\pi^+\pi^0}=
\frac{2}{3}\times\mathcal B_{K_1\to K\rho}+
\frac{4}{9}\times\mathcal B_{K_1\to K^{*}(892)\pi}+
\frac{4}{9}\times0.93\times \mathcal B_{K_1\to K^{*}_0(1430)\pi},$
where $K_1$ denotes $\bar K_1(1270)^0$.
\end{thebibliography}
\end{document}